# Guiding Trojan light beams via Lagrange points

Haokun Luo[1,†], Yunxuan Wei[1,†], Fan O. Wu[2], Georgios G. Pyrialakos[1,2], Demetrios N. Christodoulides[1,3*], Mercedeh Khajavikhan[1,3*]

[1]Ming Hsieh Department of Electrical and Computer Engineering, University of Southern California, CA 90089, USA

[2]CREOL, The College of Optics and Photonics, University of Central Florida, Orlando, FL 32816, USA

[3]Department of Physics and Astronomy, University of Southern California, CA 90089, USA

*Corresponding authors: khajavik@usc.edu (M.K.), Demetri@usc.edu (D.N.C.)

†These authors contributed equally to this work

**Abstract**:

**The guided transmission of optical waves is critical for light-based applications in modern communication, information processing, and energy generation systems. Traditionally, guiding light waves in structures like optical fibres is predominantly achieved through the use of total internal reflection. In periodic platforms, a variety of other physical mechanisms can also be deployed to transport optical waves. However, transversely confining light in fully dielectric, non-periodic and passive configurations remains a challenge in situations where total internal reflection is not supported. Here, we present an approach to trapping light that utilizes the exotic features of Lagrange points — a special class of equilibrium positions akin to those responsible for capturing trojan asteroids in celestial mechanics. This is achieved in twisted arrangements in which optical Coriolis forces induce guiding channels even at locations where the refractive index landscape is defocusing or entirely unremarkable. These findings may have implications beyond standard optical waveguiding schemes and could also apply to other physical systems such as acoustics, electron beams and ultracold atoms.**

**Main Text:**

Molding the flow of light is nowadays one of the cornerstones of modern photonic technologies[1-6]. In this respect, recent years have witnessed a flurry of intense activities on several research fronts like those associated with topological[7-10] and non-Hermitian optics[10,11], photonic crystals[2-6,12], optical metamaterials[13,14] and metasurfaces[15,16]. Of crucial importance are methodologies by means of which light can be transmitted in a guided format. Perhaps, the most prominent of them is that offered by optical fibers that are known to rely on the process of total internal reflection[1]. Unlike in fibers where lightwaves are trapped in a high-index region, in periodic photonic settings, waveguiding can also be attained through other mechanisms such as that of Bragg reflection[2-6,12,17,18], evanescent coupling[19] and bound state in the continuum effects[20]. Similarly, light can propagate in a confined manner in plasmonic[21] or active arrangements that allow for gain guidance[22]. At this point, the following question naturally arises: will it be possible to guide light in a fully dielectric and passive material system without making use of either total internal reflection or periodicity? In a wider context, can guiding occur in an open or bulk space environment, where trapping is achieved through a remote mechanism that can act and be manipulated from afar? If so, a new paradigm is established through which not only light, but also other types of waves such as acoustic or electron beams can be trapped by utilizing altogether new strategies.

In celestial mechanics, Lagrange points represent equilibrium positions whereby the gravitational attraction from two orbiting massive bodies balances the centrifugal force[23,24]. In this situation, one can identify five Lagrange points designated as $L_1$, $L_2$, …, $L_5$ with the first three ($L_1$, $L_2$ and $L_3$) being by nature unstable, while the remaining two ($L_4$ and $L_5$) are stable[23]. As a result, a small mass can be forever trapped in the vicinity of these two latter points, like for example the Trojan asteroids in the Sun-Jupiter system (Fig. 1a). Whereas at a first glance, the two-dimensional potential distribution within the co-rotating frame appears to be unstable around the $L_{4,5}$ points shown in Fig. 1a (given that is concave downwards as shown in Fig. 1a), counterintuitively, the motion of a third body can be dynamically stabilized because of the Coriolis force. In essence, as this body starts to roll down from the $L_{4,5}$ potential hill, the Coriolis force tends to continuously pull it back, thus resulting into a bounded orbit (inset in Fig. 1a). In recent years, the analogue of this process has also been used in atomic physics to produce non-dispersive wavepackets[25-27]—a prospect enabled by the isomorphism between the Coulomb force and the gravitational attraction.

In this work, we demonstrate a new approach for guiding light in transparent dielectric systems. This is achieved by exploiting the intriguing characteristics of the Lagrange points and the equivalence between the paraxial wave equation for light and the non-relativistic Schrödinger equation. The resulting optical Trojan beams can be captured even in defocusing environments at points where the refractive index profile (within the stationary frame) is totally ordinary—with no features that could foretell a guiding behavior. We show that, in the optical domain, this mechanism can be established in a versatile manner through a variety of attractive/repulsive index potentials. Given that Newton's third law governing the dynamics of celestial bodies is in our optical setup inconsequential in establishing a rotating potential, multiple Lagrange points can now be simultaneously generated at will in "many-body" configurations, an aspect that is impossible in other settings.

Figure 1b depicts a schematic of the experimental arrangement used to observe optical Trojan beams. These beams are excited in ~30 cm-long glass cylinders with a radius of $b = 1.18$ cm that are filled with cured polydimethylsiloxane (PDMS) of refractive index $n_0 = 1.46$. To produce optical Lagrange points, we exploit the large thermo-optic effect offered by PDMS ($dn/dT = -4.5 \times 10^{-4}$ K$^{-1}$). This is achieved by inserting a twisted iron wire (of radius $a = 275$ μm) in the tube before the PDMS is cured. Electric current is then passed through this wire that in turn heats the PDMS in order to induce a logarithmic defocusing spiral index potential (Supplementary, section 1,2) through the thermo-optic process (Fig. 2a).

In addressing the prospect of trapping light at a Lagrange point, let us consider, in the stationary coordinate system $(x, y)$, the logarithmic defocusing index profile depicted in Fig. 2a, which is typical to that thermally produced in our experiments. This long-range potential is spiraling around the center $C$ with a pitch of $\Lambda = 8$ cm. In this example, the index contrast induced (with respect to the surface of the metallic wire) is approximately $2.25 \times 10^{-2}$. On the other hand, the same index potential remains static, when viewed within the co-rotating frame, which is shown in Fig. 2b along with the corresponding iso-contour lines. In this frame, because of centrifugal effects, the effective potential is modified, leading to two Lagrange points (designated as $L_A$ and $L_B$) where the optical attraction or repulsion is balanced. In this case, it so happens that only $L_A$ is stable while $L_B$ is unstable. The position of $L_A$ is also marked in Fig. 2a for clarity. To showcase the trapping potential of the Lagrange points, we begin our analysis by examining ray dynamics in the vicinity of $L_A$ (Supplementary, section 3). Fig. 2c shows the calculated ray trajectories, from where one can conclude that light is captured around this Lagrange point. As in the case of Trojan asteroids, here stability is introduced because of Coriolis forces. Interestingly, this is possible even though the potential is defocusing (hence, repulsive) and it is produced only by one source, unlike the Sun-Jupiter system[23]. As indicated before, the index landscape around the $L_A$ point, when viewed within the stationary $(x, y)$ system (Fig. 2a), happens to be completely ordinary, with no special features that could ever foretell that light can be trapped. The stability criteria for these optical Lagrange points are provided in the supplementary, section 4.

While the ray dynamics do suggest that light can be guided at a Lagrange point, it is still imperative to formally assert this possibility within the framework of wave optics. To do so, we use the paraxial wave equation, a reduced form of the Helmholtz electromagnetic problem in weakly guiding arrangements, that so happens to be isomorphic to the Schrödinger equation. In normalized units, this is given by $i\partial_z \psi = \widehat{H} \psi$ where $\widehat{H} = -(\partial_{xx} + \partial_{yy})/2 + V(x, y, z)$ is the Hamiltonian operator and $\psi$ is the slowly varying electric field amplitude. The potential $V(x, y, z)$ that appears in the Hamiltonian is directly proportional to the refractive index profile (Supplementary, section 5). The paraxial equation can be more conveniently studied within the co-rotating frame $(u, v, \xi)$ once a "magnetic" vector potential $\boldsymbol{A}$ is introduced via $\boldsymbol{A} = \boldsymbol{\Omega} \times \boldsymbol{r}$ to account for optical Coriolis effects[8,28], where $\Omega = 2\pi/\Lambda$ is the rotation rate. In this case, by keeping in mind $\xi = z$, the optical beam evolution equation takes the form:

$$i\frac{\partial \psi}{\partial z} = \left[\frac{1}{2}(\boldsymbol{p} - \boldsymbol{A})^2 + V_{\Omega,\text{eff}}(u, v)\right] \psi, \tag{1}$$

where $\boldsymbol{p} = -i\hat{u}\partial/\partial u - i\hat{v}\partial/\partial v$, and $V_{\Omega,\text{eff}}(u, v) = V(u, v) - \Omega^2 r^2/2$ (Supplementary, section 6). Equation (1) is solved numerically to identify bound modes in the form of $\psi = e^{i\sigma z} R(u, v) e^{i\Psi(u,v)}$, centered at a stable Lagrange point where $\sigma$ denotes their corresponding eigenvalue. Figure 2d depicts the intensity profile of the ground state associated with the $L_A$ point in Fig. 2b. In this case, the intensity of the fundamental Trojan mode is elliptical, as one could expect given that the index potential around $L_A$ has approximately an elliptic paraboloidal dependence. The propagation characteristics of the Trojan beam are then numerically investigated within the stationary $(x, y)$ frame (Fig. 2e). Evidently, being a mode, the Trojan beam remains invariant during propagation while twisting around the center point $C$. In other words, because of Coriolis forces, the beam is stably trapped around the Lagrange point, thus overcoming any diffraction and attraction/repulsion effects. These results are in accord with those obtained from ray dynamics. The correspondence between wave and ray optics can be directly established through the Ehrenfest theorem (Supplementary, section 7).

To observe Trojan beams, experiments are conducted in PDMS filled cylinders (Fig. 1b). The optical dynamics of these beams are monitored both at the output facet of the cylinder as well as from the side using an imaging camera. To enable sideview observations, the PDMS is mixed with a small amount of $TiO_2$ nanoparticles (165 nm in diameter). In this experiment, the spiral iron wire is positioned at 1.85 mm from the center $C$, having a pitch $\Lambda = 8$ cm. The Trojan mode is then excited by a Gaussian beam with a spot size of $w_0 \simeq 110$ μm from a He-Ne laser ($\lambda_0 = 632.8$ nm), right at the Lagrange point of this arrangement. A DC current $I = 4.0$ A is passed through the wire to heat the sample. In this case, the surface of the wire is estimated to be at 50 K above the outer surface temperature of the cylinder, and the resulting temperature distribution leads to a spiraling logarithmic index potential $\Delta n \simeq -2.25 \times 10^{-2}[1 - \ln(\rho/a)/\ln(b/a)]$ where $\rho$ is the distance from the wire center (Supplementary, section 2). In this configuration, the Lagrange point $L_A$ is positioned $\sim 430$ μm away from the center $C$. Experiments demonstrating that the Trojan beam is captured around $L_A$ are shown in Fig. 3a,b, demonstrating that indeed the beam retains its spot size both at the output facet (Fig. 3b) and during propagation (Fig. 3b). At the output, the mean value of the spot size radius is $\sim 112$ μm with an ellipticity of 0.49. The variation of the average beam spot size as a function of distance and current is depicted in Fig. 3c,d. Meanwhile, if the current is turned off, the beam is no longer trapped and finally expands after $\sim 30$ cm of propagation to $w \simeq 385$ μm because of diffraction (Fig. 3e,f). We next keep the wire straight while conveying the same current $I = 4.0$ A. Interestingly, in this scenario, not only the beam diffracts but is also strongly repelled by the defocusing (repulsive) index potential and forced into a self-bouncing trajectory at the boundary of the cylinder (Fig. 3g). This response confirms that the Lagrange point is responsible for the behavior displayed in Fig. 3a,b.

To demonstrate in a definitive manner that what is observed in our setup is a Trojan beam, we experimentally detect its phase structure that so happens to be quite unique to this trapped state. To explain the origin and importance of this phase $\Psi(u, v)$, let us approximately represent the effective potential landscape in the vicinity of a Lagrange point $(u_i, v_i)$ where it is maximum via a Taylor series. In this case, one can obtain closed form solutions for the bound states of Eq. (1) (Supplementary section 8), from which we find that $\Psi(u, v) = \alpha u + \beta v + \gamma u v$. The first two terms $(\alpha, \beta)$ are associated with the tilt of the Trojan beam when traversing its helical path while the last term $\gamma u v$ governs the internal energy flow that allows this elliptical state to reorient itself during propagation. We would like to emphasize that the $\gamma u v$ phase is special to these Trojan entities, given that it doesn't appear in standard dielectric waveguide settings or gain-guiding systems. Numerical simulations carried out for the index profiles used in our experiments (Fig. 2b), lead to similar results (Fig. 3h). To detect the $\gamma u v$ phase, we perform interferometric measurements using a Mach-Zehnder arrangement where the Trojan beam in one of the arms is appropriately tilted so as to remove the effects from the $(\alpha u + \beta v)$ contributions. The resulting X-shaped interferogram (produced by the $\gamma u v$ term) is shown in Fig. 3i. This provides irrefutable evidence that what is observed in our setup is a Trojan beam. In optical arrangements, it is possible to excite multiple Lagrange points from several potential sources, something that is not feasible in celestial mechanics because many-body systems are chaotic (Supplementary section 9). We would like to emphasize that what is discussed here is fundamentally different from results previously obtained in transversely periodic arrangements whereby diffraction effects can be arrested even in defocusing potentials by exploiting the effective diffraction properties (effective mass) of Bloch modes within the Brillouin zone[29,30]. On the other hand, in bulk media where the magnitude and sign of the diffraction cannot be altered, there is no other mechanism known (apart from the one presented in this work) by means of which light trapping is possible even in defocusing environments.

We next conduct experiments in a double-helix wire configuration when embedded in a cured PDMS cylinder. In this case, the Lagrange point is located right at the center $C$. Each one of the two spiral wires is positioned at 1.8 mm from the center $C$ and the helix pitch used is $\Lambda = 6.3$ cm. When no current is flowing through the system, the optical beam diffracts to a spot size $w \simeq 375$ μm after ~30 cm of propagation (Fig. 4a,b). On the other hand, once a current $I = 3.5$ A is passed through the two wires (connected in series), an elliptical Trojan state is formed when excited with a Gaussian beam of spot size $w_0 \simeq 110$ μm (Fig. 4c,d). Our results clearly indicate that the beam is trapped around the Lagrange point given that at the output its mean spot size radius is approximately 120 μm with an ellipticity of 0.9 (Fig. 4d). The dependence of the average beam spot size on distance is depicted in Fig. 4e. The X-shaped interferogram resulting from the corresponding $\gamma uv$ phase is also depicted in Fig. 4f. Experiments were also performed when the two wires were kept straight, in which case no Lagrange point is produced, Fig. 4g. To some extent, trapping light in this double-helix arrangement is reminiscent of charged particle confinement in Paul traps[31,32] that is achieved via parametric interactions – a stabilization process that is nonetheless different from the one mediated by the Coriolis force.

In conclusion, we have demonstrated for the first time a new methodology for trapping light by utilizing the unique features of Lagrange points. The resulting Trojan beams can be guided even in defocusing refractive index landscapes because of optical Coriolis effects. This is enabled through their unique phase distribution that allows their internal energy to be appropriately rearranged along their trajectory. Our work may open new avenues in guiding optical waves in settings where traditional approaches are not possible, as for example, in liquid environments. Of interest would be to investigate the possibility of observing these Trojan states in amplifying configurations where optical gain can exert long-range attractive or repulsive forces on a signal that is not confined in a standard dielectric waveguide. The prospect of guiding and deflecting light at Lagrange points induced by orbiting ultra-massive bodies like black holes or neutron stars can be another exciting direction in astrophysics.

## Acknowledgements

This work was supported by the Air Force Office of Scientific Research (AFOSR) Multidisciplinary University Research Initiative (MURI)award on Novel light-matter interactions in topologically non-trivial Weyl semimetal structures and systems (award no. FA9550-20-1-0322)(M.K., D.N.C., H.L., Y.W., F.O.W. and G.G.P.), AFOSR MURI award on Programmable systems with non-Hermitian quantum dynamics(award no. FA9550-21-1-0202) (M.K., D.N.C., H.L., Y.W., F.O.W. and G.G.P.), ONR MURI award on the classical entanglement of light (award no. N00014-20-1-2789) (M.K., D.N.C., H.L., Y.W., F.O.W. and G.G.P.), AFRL – Applied Research Solutions (S03015) (FA8650-19-C-1692) (M.K.), W.M. Keck Foundation (D.N.C.), MPS Simons collaboration (Simons grant no. 733682) (D.N.C.) and US Air Force Research Laboratory (FA86511820019) (D.N.C.).

## Author Contributions

D.N.C. and M.K. conceived the idea. H.L., Y.W., F.O.W. and G. G. P. developed the theory. H.L and Y.W conducted the simulations, data analysis, and the experiments. All the authors contributed to the writing of the original draft, review, and editing.

## Competing interests

The authors declare no competing interests.

# Figure Legends/Captions (for main text figures)

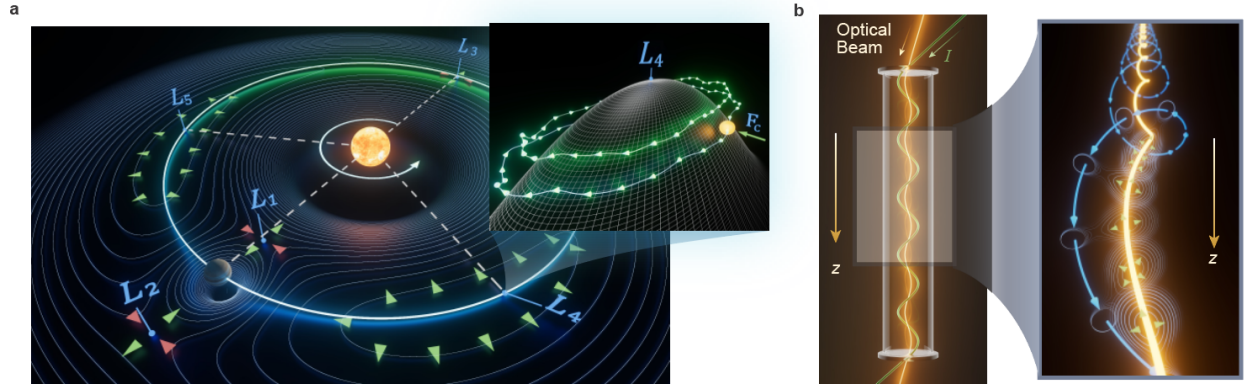

**Fig. 1|Celestial and optical beam dynamics in the vicinity of a stable Lagrange point. a,** Lagrange points in the Sun-Jupiter system. In the co-rotating frame, the potentials associated with the three unstable colinear Lagrange points ($L_1$, $L_2$, $L_3$) are saddle-shaped while those of $L_4$ and $L_5$ are stable being maxima. The inset in (**a**) shows the stable trajectory of an asteroid, when captured around an $L_4$ Lagrange point because of the Coriolis force. **b,** Experimental setup used to observe optical Trojan bound states (bright yellow beam). A stable Lagrange point is established via the thermo-optic effect by passing current $I$ through a helical iron wire (shown in green) embedded in a cured PDMS cylinder. The inset in (**b**) displays in a schematic manner the trapping of a Trojan beam (depicted in bright yellow) at a Lagrange point induced by a helicoidal index potential (blue line).

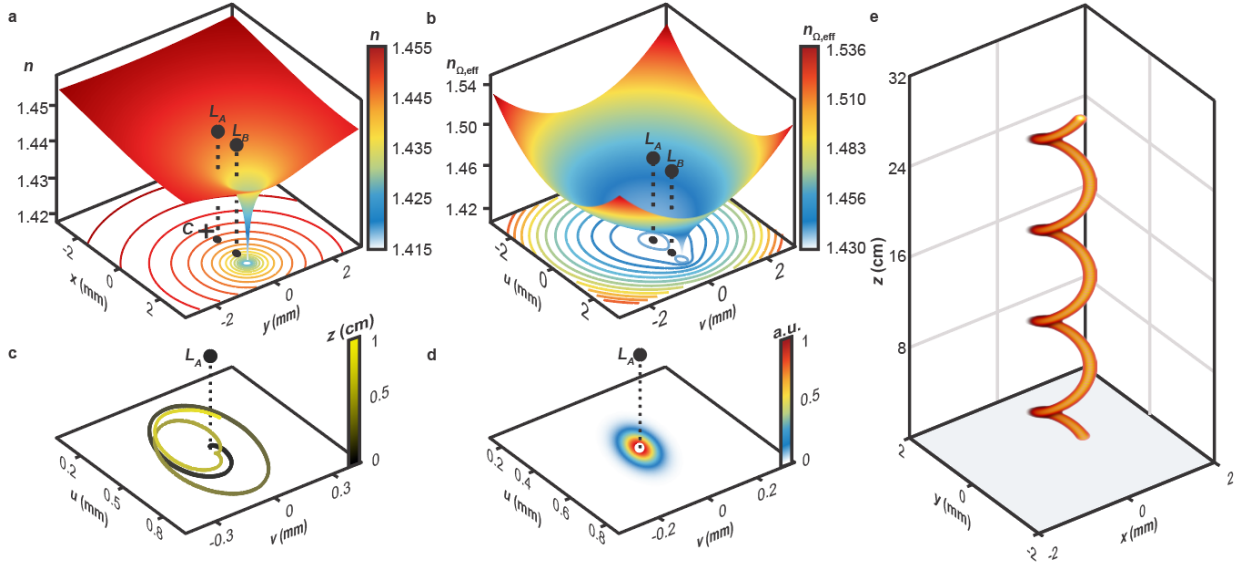

**Fig. 2|Light propagation dynamics around a stable Lagrange point. a,** An induced logarithmic defocusing spiraling index potential when viewed within the stationary frame $(x, y)$. This index profile rotates along $z$ at a constant angular velocity $\Omega$ around the center $C$. The corresponding iso-contour lines are also shown. **b,** In the co-rotating frame $(u, v)$, the effective index potential now involves centrifugal effects and exhibits two Lagrange points, $L_A$ and $L_B$. $L_B$ is a saddle point and hence is unstable. On the other hand, $L_A$ exhibits a minimum where the dynamics can be stabilized through the optical Coriolis force. Note that in optics, the effective index distribution in the co-rotating coordinate system is $\Delta n_{\Omega,\text{eff}} \propto -V_{\Omega,\text{eff}}$ (Supplementary section 4). For comparison, the positions of $L_A$ and $L_B$ are also marked in (**a**). **c,** Stable ray dynamics unfolding around $L_A$ in the index landscape depicted in (**a**) and (**b**) as viewed within in the co-rotating frame. **d,** Numerically obtained normalized intensity distribution in arbitrary units (a.u.) for the fundamental Trojan optical mode centered at $L_A$, corresponding to (**a**) and (**b**). In this case, the mode is elliptical in the $(u, v)$ system. **e,** Stable propagation of the optical Trojan ground state shown in (**d**), as obtained from numerically solving Eq. (1) along $z$. The beam remains invariant along its helical path. The helix pitch in (**a**)-(**e**) is $\Lambda = 8$ cm.

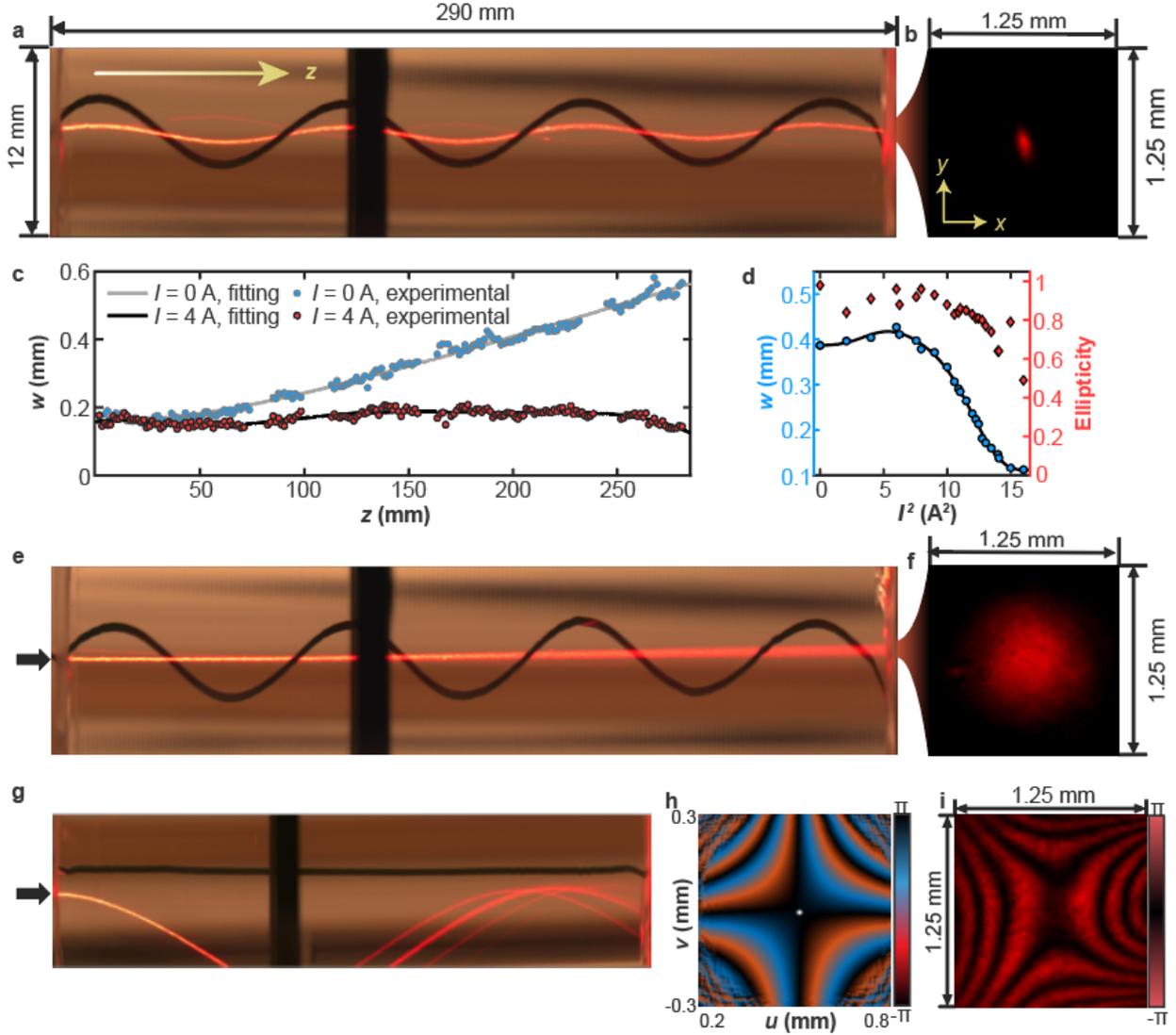

**Fig. 3|Trojan beam guiding in the index potential produced by a single helical heat source. a,** Side-view image of the trapped fundamental Trojan mode (red beam) when helicoidally traversing a PDMS filled tube. The spiraling wire, carrying a current $I = 4.0$ A, is also shown in the background. The spot size of the Gaussian beam at the input is ~110 µm. **b,** Intensity profile of the Trojan mode at the output facet, after a distance approximately 30 cm. The mean value of the spot size radius is ~112 µm with an ellipticity of 0.49 (minor/major axis). **c,** Variation of the beam's mean spot size as a function of distance for $I = 4.0$ A. **d,** Dependence of the Trojan mode's output mean spot size and ellipticity versus current $I^2$. **e,** Side-view image of the diffraction dynamics of the input Gaussian beam when the current is turned off. **f,** For the case shown in (**e**), the circular beam diffracts to a spot size of ~385 µm. **g,** When the wire is kept straight while carrying a current $I = 4.0$ A, the light beam is strongly repelled by the defocusing index landscape, leading to a self-bouncing behavior at the surface of the cylinder. **h,** Numerically obtained phase structure associated with the fundamental Trojan mode centered around $L_A$ (depicted as a white dot). The presence of the characteristic X-shaped $\gamma uv$ term is evident. **i,** Experimentally observed wavefront phase of the Trojan state corresponding to (**a**) and (**b**) detected via interferometry. The aspect ratio of (**a**) and (**e**) is 4.5:1 while the one of (**g**) is 3.2:1. The dark stripe shown in (**a**), (**e**) and (**g**) arises from the tube holder.

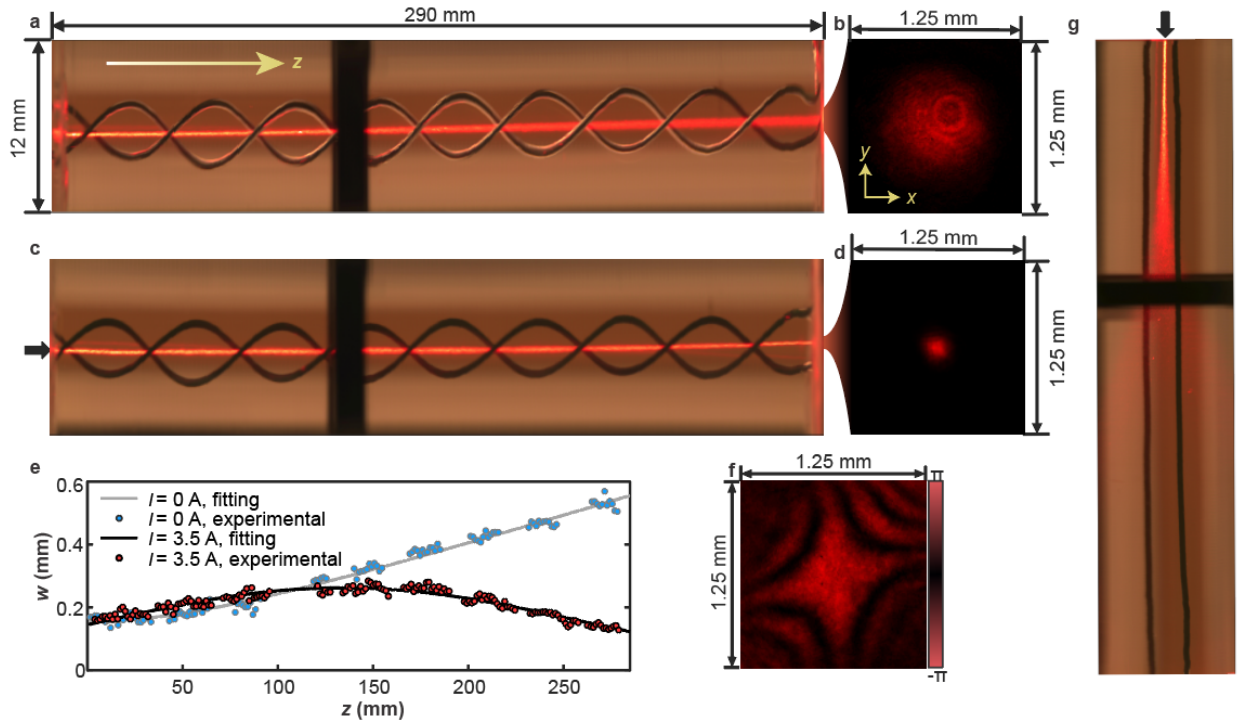

**Fig. 4|Trojan beam trapping in the index landscape produced by a double-helix current source. a,** A Gaussian beam with a spot size of ~110 μm is injected into the sample when the current is $I = 0$ A. **b,** After ~30 cm of propagation the beam diffracts to a spot size of 375μm. **c,** A Trojan optical mode is established when the current $I = 3.5$ A. In this case, the Lagrange point is located at the center of this double-helix wire system ($\Lambda = 6.3$ cm). **d,** Output intensity profile of this Trojan state having a mean spot size of ~120 μm with an ellipticity of 0.9. **e,** Variation of the beam's mean spot size as a function of distance for $I = 3.5$ A. **f,** Interferometrically observed phase structure of the trapped mode shown in (**c**) and (**d**), indicative of the X-shaped phase term. **g,** No Lagrange point is induced when the wires are kept straight and therefore the beam diffracts. The aspect ratio in (**a**), (**c**) and (**g**) is 4.5:1. The dark stripe shown in (**a**), (**c**) and (**g**) arises from the tube holder.

## Methods

**Experimental optical setup**

The experimental setup depicted in Extended Data Fig. 1, comprises three main characterization components. These include (1) beam excitation, (2) sideview imaging and (3) bottom view imaging and interference. For exciting a beam, a linearly polarized HeNe laser (Thorlabs HNL050LB) operating at a wavelength of 632.8 nm is used. Its output Gaussian beam is 5× expanded through a 4$f$ system and then is focused to a spot size of $w_0 \simeq 110$ µm through a convex focusing lens $C_f$ having a focal length of $F_{C_f}$. The position of the focal spot as well as the launching angle is carefully adjusted so as the input Gaussian beam couples efficiently to the Trojan beam near the upper surface of the glass cylinder. A DC current source is connected to the helical iron wire embedded in the cured PDMS, for heating and generating the required logarithmic twisted refractive index potential. To acquire a sideview of the beam dynamics inside the tube, a CMOS camera (EO-1312c, Edmund Inc.) is used in conjunction with a lens, attached to a vertically movable stage along the propagation direction. A microscope, involving a 10× objective ($NA = 0.28$) and a $F = 125$ mm lens, is employed to image the transverse profile of the Trojan beam at the bottom facet. Interferometric measurements are performed by splitting the light after the microscope into two beams, the first one passing through a 2.86× 4$f$ system to expand it into a flat wavefront while the other one is appropriately attenuated to perform the interferometric measurement. A linear polarizer is inserted into the reference arm to adjust the interference contrast. The interference pattern is finally projected with a magnification 0.5× onto a CMOS camera through a 4$f$ system. The spatial frequency spectra of the two interfering beams are matched at the focal plane of the $F = 200$ mm lens in this 4$f$ system so that the tilt phase shift terms ($\alpha, \beta$) are eliminated by tilting the reference beam. In imaging the optical beam profiles at the output facet (collected by a CMOS camera BladeCam-HR, Dataray Inc.), both the reference arm and the attenuator are removed. In our cases, no polarization rotation was observed given that the geometric phase was insignificant for the small helix radius used.

In characterizing the phase structure of the Trojan beams, the focal length of the $C_f$ lens is changed depending on the current $I$ used. This is done to observe the phase structure by expanding the intensity distribution that further facilitates the interferometric measurements. Extended Data Table 1 provides the focal length of $C_f$ for different currents $I$. (The focal length of $C_f$ for different currents $I$ are provided in Supplementary Information section S1.)

Apart from the camera lens providing ~2.5×, the magnification of the sideview system is also affected by the circular interface between air and the PDMS filled glass tube. Given that the refractive indices of glass ($n_{\text{glass}} = 1.457$) and PDMS ($n_0$ ~1.46) are approximately the same, then under paraxial conditions, the corresponding object-virtual image relation follows the equation

$$\frac{n_0}{s_0} - \frac{n_{\text{air}}}{s_i} = \frac{n_0 - n_{\text{air}}}{b_{\text{out}}}, \qquad (2)$$

where $s_0$ and $s_i$ represent the distance from the interface associated with the object and the image as measured inside the cylinder, respectively. Here, $n_{\text{air}} = 1$, and $b_{\text{outer}} = 1.50$ cm is the outer radius of the glass cylinder. The resulting magnification is given by

$$M = \frac{1}{1 - \left(1 - \frac{n_{\text{air}}}{n_0}\right)\frac{s_0}{b_{\text{outer}}}}. \tag{3}$$

Suppose that the object is placed at the center of the tube such that $s_0 = b_{\text{outer}}$, the magnification is exactly

$$M = \frac{n_0}{n_{\text{air}}} = 1.46. \tag{4}$$

This inherent magnification is taken into account when estimating the beam spot size from sideview imaging.

**Sample preparation**

Our sample involves a spiraling iron wire surrounded by cured PDMS in a ~30 cm long glass tube. The tube is covered by two transparent acrylic disks so as to avoid dealing with curved surfaces arising from surface tension and any other distortion effects from thermal expansion. The radius of the iron wire is $a = 0.275$ mm, while the outer and inner radii of the glass tube are $b_{\text{outer}} = 15.0$ mm and $b = 11.8$ mm, respectively. To image the optical beams from the side, a small amount of TiO$_2$ nanoparticles (165 nm in diameter, Nanomaterials Inc) are uniformly distributed in PDMS (Microlubrol Sylcap 284-F) as scatterers. (Typical samples used in single- and double-helix wire configurations are provided in Supplementary Information section S1.) Extended Data Fig.2 shows typical samples used in single- and double-helix wire configurations.

The cured PDMS is prepared by using the following steps. First, the liquid PDMS and a small amount of TiO$_2$ nanoparticles are mixed together by utilizing a magnetic stirrer at a temperature of 140 °C for 12 hours. It is then allowed to cool to room temperature and eventually is mixed with a curing agent (Sylcap 284-F: Polydimethylsiloxane, vinyldimethylsiloxy terminated–68083-19-2. Methylhydrogen Polysiloxane–68988-57-8.) at a volume ratio of 10:1. After adequately stirred, the mixture is put into a vacuum chamber for 40 minutes to avoid any bubble generation during the curing process. Finally, the liquid PDMS is poured into the glass tube and let it stand for more than 18 hours at room temperature to get fully cured.

**Ray stability analysis around a Lagrange point**

The stability of ray dynamics around a stable Lagrange point located at $(u_i, v_i)$ is analyzed. In the co-rotating $(u, v)$ system, the potential landscape in the vicinity of this point is locally a maximum, that is, to first order, its Taylor series is given by $V_{\Omega,\text{eff}} = -\Delta n_{\Omega,\text{eff}}(u,v)/n_0 = V_{\max} - (\omega_1^2(u - u_i)^2 + \omega_2^2(v - v_i)^2)/2$. If we let $\tilde{u} = u - u_i, \tilde{v} = v - v_i$, the equations governing the ray dynamics (Supplementary Information Section 4) is given by

$$\ddot{\tilde{u}} = 2\Omega\dot{\tilde{v}} + \omega_1^2 \tilde{u}, \tag{5-1}$$

$$\ddot{\tilde{v}} = -2\Omega\dot{\tilde{u}} + \omega_2^2 \tilde{v}. \tag{5-2}$$

By solving the corresponding eigenvalue problem (Supplementary Information Section 4), one finds that the ray trajectories around the Lagrange point are stable only when

$$2\Omega > \omega_1 + \omega_2. \tag{6}$$

Eq. (6) provides the stability criterion required to trap an optical ray around a stable Lagrange point.

**Analytical solution for the fundamental Trojan mode in a twisted parabolic potential**

Here we provide an analytical solution for the fundamental Trojan mode in a twisted parabolic elliptical potential. The Lagrange point is located at $(u_i, v_i)$ within the $(u, v)$ system. To do so, we first obtain this solution when the potential is shifted to the center, in which case, to first order $V_{\Omega,\text{eff}} = V_{\max} - (\omega_1^2 u^2 + \omega_2^2 v^2)/2$. In this case, the optical envelope obeys

$$i\frac{\partial \psi}{\partial z} = -\frac{1}{2}\left(\frac{\partial^2 \psi}{\partial u^2} + \frac{\partial^2 \psi}{\partial v^2}\right) + i\Omega\left(u\frac{\partial \psi}{\partial v} - v\frac{\partial \psi}{\partial u}\right) + \frac{1}{2}\Omega^2(u^2 + v^2)\psi + \left(V_{\max} - \frac{(\omega_1^2 u^2 + \omega_2^2 v^2)}{2}\right)\psi. \tag{7}$$

The ground state of Eq. (7) has the form of an elliptical Gaussian function[33]

$$\psi = Ne^{-pu^2}e^{-qv^2}e^{i\gamma uv}e^{i\sigma z}e^{-iV_{\max}z}, \tag{8}$$

where $p, q > 0$ ($p, q \in \mathcal{R}^+$) and $N$ is a normalization factor. The parameters involved in this solution are given by (Supplementary Information Section 8)

$$\gamma = \frac{\omega_2 - \omega_1}{\omega_2 + \omega_1}\Omega, \tag{9}$$

$$p = \frac{[4\Omega^2 - (\omega_1 + \omega_2)^2]^{\frac{1}{2}}}{2(\omega_1 + \omega_2)}\omega_1, \tag{10}$$

$$q = \frac{[4\Omega^2 - (\omega_1 + \omega_2)^2]^{\frac{1}{2}}}{2(\omega_1 + \omega_2)}\omega_2. \tag{11}$$

$$\sigma = -\frac{[4\Omega^2 - (\omega_1 + \omega_2)^2]^{\frac{1}{2}}}{2}. \tag{12}$$

Note that this solution is only possible if $2\Omega > \omega_1 + \omega_2$, which is identical with Eq. (6). This solution can be translated to $(u_i, v_i)$ using a gauge phase[34], i.e., $\psi_{u_i,v_i} = \psi(u - u_i, v - v_i)e^{i\Phi(u,v)}$, where $\Phi(u, v) = \Omega(u_i v - v_i u)$.

The dynamics of the Trojan mode can be solved numerically under any arbitrary initial conditions using a beam propagation method that relies on fast Fourier transforms (BPM-FFT). The Trojan modes supported by the actual effective potential around a stable Lagrange point (like the one depicted in Fig. 2b) are obtained by numerically solving Eq. (1). In this case, the eigenvalue problem is solved using finite-difference methods (FDM).

## Data availability

Source data are available for this paper. All other data that support the plots within this paper and other findings of this study are available from the corresponding author upon reasonable request.

## Code availability

The used numerical codes are based upon MATLAB and COMSOL and are available upon reasonable request to the corresponding authors.

## Methods-only References

**33.** Rebane, T. K. Two-dimensional oscillator in a magnetic field. *J. Exp. Theor. Phys.* **114**, 220-225 (2012).

**34.** Landau, L. D. & Lifshitz, E. M. in *Quantum Mechanics (Third Edition)* 453-471 (Pergamon, 1977).

## Inclusion & Ethics

All authors acknowledge the Global Research Code on the development, implementation and communication of this research. For the purpose of transparency, we have included this statement on inclusion and ethics. This work cites a comprehensive list of research from around the world on related topics.

## Additional information

**Supplementary Information** is available for this paper.

**Correspondence and requests for materials** should be addressed to M.K. or D.N.C.

**Reprints and permissions information** is available at www.nature.com/reprints.